\documentclass[sigconf]{acmart}
\copyrightyear{2026}
\acmYear{2026}
\setcopyright{cc}
\setcctype{by}
\acmConference[EASE '26]{International Conference on Evaluation and Assessment in Software Engineering}{June 09--12, 2026}{Glasgow, United Kingdom}
\acmBooktitle{International Conference on Evaluation and Assessment in Software Engineering (EASE '26), June 09--12, 2026, Glasgow, United Kingdom}
\acmDOI{10.1145/3816483.3816551}
\acmISBN{979-8-4007-2348-3/2026/06}

\usepackage{float}
\usepackage{subcaption}
\usepackage{tabularx}
\usepackage{todonotes}
\usepackage{tcolorbox}
\usepackage{amsmath}

\usepackage{amssymb}
\usepackage{balance}
\usepackage{booktabs}
\usepackage{caption}
\usepackage{colortbl}
\usepackage{xcolor}
\usepackage{enumitem}
\usepackage{comment}
\usepackage{soul}
\usepackage{balance}

\begin{document}

\title{Comparing Smart Contract Paradigms: A Preliminary Study of Security and Developer Experience}

\author{Matteo Vaccargiu}
\affiliation{%
  \institution{University of Cagliari}
  \city{Cagliari}
  \country{Italy}
}
\email{matteo.vaccargiu@unica.it}

\author{Andrea Pinna}
\affiliation{%
  \institution{University of Cagliari}
  \city{Cagliari}
  \country{Italy}
}
\email{andrea.pinna83@unica.it}

\author{Maria Ilaria Lunesu}
\affiliation{%
  \institution{University of Cagliari}
  \city{Cagliari}
  \country{Italy}
}
\email{mariai.lunesu@unica.it}

\author{Giuseppe Destefanis}
\affiliation{%
  \institution{University College London}
  \city{London}
  \country{United Kingdom}
}
\email{g.destefanis@ucl.ac.uk}

\settopmatter{authorsperrow=4}

\begin{abstract}
Smart contract vulnerabilities have caused billions in financial losses, motivating the question of whether resource-oriented languages like Move, which encode safety guarantees in type systems, can reduce the security overhead imposed by imperative languages like Solidity. We present a preliminary mixed-methods study analyzing 12 functionally-equivalent contract pairs implemented in both Solidity and Move by the same development team, complemented by a survey of 11 developers experienced in both languages. Quantitative analysis reveals that Move reduces explicit security overhead by 60\% ($p=0.002$, Cohen's $d=-1.75$) at the cost of 47\% larger code size ($p=0.002$, $d=1.90$), while maintaining identical cyclomatic complexity. Developer surveys show moderate learning difficulty but higher safety confidence in Move (Median=6/7, 10 of 11 above neutral), with 55\% preferring Move for security-critical applications despite ecosystem maturity gaps. 
These preliminary findings suggest resource-oriented paradigms shift security from runtime validation to compile-time guarantees, providing initial evidence for paradigm effects on smart contract development, though adoption requires investment in learning and tooling.
\end{abstract}

\maketitle

\section{Introduction}

Programming language design influences software quality and security. Language features shape how developers express logic, manage complexity, and prevent errors~\cite{chen2021empirical}. Some languages rely on developers to manually implement safety checks and validation logic, while others encode guarantees directly in type systems or runtime environments. These design choices affect code correctness and development productivity, maintainability, and the cognitive burden developers face when reasoning about program behavior. Understanding and quantifying these differences enables developers and organizations to make evidence-based decisions about language adoption and training investments.

These concerns become even more important in blockchain programming, where smart contracts manage billions of dollars in digital assets without intermediaries. Unlike traditional software, smart contracts cannot be updated after deployment. This permanence creates a critical security challenge: once deployed, vulnerabilities become permanent with no possibility of patching~\cite{perez2021smart} unless specific patterns are used. 
High-impact attacks have demonstrated the severity of this constraint, with DeFi protocols alone experiencing an estimated \$6.45 billion in financial damage~\cite{zhou2023sok}. A single reentrancy bug caused the infamous DAO hack, resulting in \$60 million in losses~\cite{atzei2017survey}. 

Smart contract vulnerabilities have driven calls for specialized software engineering practices~\cite{destefanis2018smart}. Research has documented how traditional software metrics manifest differently in blockchain-oriented software~\cite{ortu2019comparing}, revealing distinct patterns in code complexity, coupling, and quality indicators. Analysis of smart contract repositories shows that vulnerability patterns correlate with specific code characteristics~\cite{ibba2024curated}, while studies of developer behavior reveal how community dynamics and emotional patterns influence code quality in projects like Solidity~\cite{vaccargiu2025more}. Despite previous research on vulnerability detection~\cite{durieux2020empirical}, we lack empirical evidence on how programming paradigms affect the security burden developers face.

Solidity, the main smart contract language for Ethereum, follows an imperative, account-based paradigm where developers manually implement security checks. 
Move, designed for the Sui, Aptos and IOTA blockchains, adopts a resource-oriented paradigm inspired by linear types, where the type system enforces ownership and prevents common vulnerabilities by construction. Recent implementations demonstrate Move's practical advantages in domains requiring secure ownership tracking, such as supply chain management~\cite{11039281}. 
This paradigm shift moves smart contract security from tool-based bug detection to language-level prevention through design. However, empirical evidence comparing these paradigms remains limited. The extent to which Move's type system reduces security overhead—versus shifting complexity to type annotations—requires empirical validation. The developer experience of this paradigm transition, including learning barriers and perceived safety benefits, is essential for language designers, organizations selecting smart contract platforms, and the broader evolution of blockchain programming languages.

To address this gap, we present a preliminary empirical comparison of imperative and resource-oriented smart contract programming through a mixed-methods study. We analyze 12 functionally-equivalent contract, each implementing identical use cases in both Solidity and Move by the same development team. We complement this code analysis with a survey of 11 developers who have implemented contracts in both languages. While limited in scale, this controlled comparison isolates paradigm effects while controlling for developer skill, coding style, and functional requirements.

We structure the study around two research questions:

\textbf{RQ1: How do imperative and resource-oriented paradigms differ in code characteristics and security overhead?} 

Prior work has focused on vulnerability detection tools~\cite{durieux2020empirical,ibba2024curated} rather than comparing how paradigms affect security overhead. We quantify this through nine metrics examining code size, complexity, explicit security checks, type annotations, and resource operations. This reveals whether Move's type system reduces security code or merely shifts complexity to type declarations.

\textbf{RQ2: How do developers experience and perceive the transition from imperative to resource-oriented smart contract programming?} 

Research on Solidity development~\cite{vaccargiu2025more} documents developer behavior patterns, but paradigm transition experiences lack systematic investigation. We address this through surveys examining perceived learning difficulty, documentation quality, and safety confidence. This reveals whether the resource-oriented paradigm's conceptual demands outweigh its perceived safety benefits—identifying specific adoption barriers in tooling, documentation, or conceptual understanding that inform training and ecosystem development.

The code to replicate the study is available at this \href{https://figshare.com/s/4e9d1d32b63aca33ccba}{\textbf{link}}.
\vspace{-6pt}
\section{Related Works}

\paragraph{Security Challenges in Smart Contracts}

Large-scale empirical studies document systemic vulnerabilities. Torres et al.~\cite{torres2018osiris} identified 42,108 integer bugs in 1.2 million contracts, Nikolic et al.~\cite{nikolic2018finding} flagged 34,200 contracts with trace vulnerabilities (including the Parity \$200M bug), and Yang et al.~\cite{yang2024uncover} detected exploitable reentrancy across hundreds of incidents. However, syntactic vulnerability does not equal exploitation~\cite{perez2021smart}. Detection tools show limitations: Durieux et al.~\cite{durieux2020empirical} revealed high false positive rates across tools. Developer behavior studies show heavy reliance on explicit security patterns despite their limitations: Mitropoulos et al.~\cite{mitropoulos2024broken} documented widespread but often incorrect use of \texttt{require}/\texttt{assert} across hundreds of thousands of contracts, while Kado et al.~\cite{kado2024empirical} and Hwang and Ryu~\cite{hwang2020gap} found that patterns like Unchecked Call persist through code cloning, leaving many contracts vulnerable despite known patches.

These studies establish that developers struggle with explicit security checks, but do not compare how paradigm choice affects baseline security overhead—a gap we address.

\vspace{-4pt}
\paragraph{Paradigms Empirical Comparisons}

Multiple blockchain platforms explore alternatives to Solidity's imperative model. Brünjes and Gabbay~\cite{brunjes2020utxo} compared account-based (Ethereum/Solidity) with UTxO-based (Cardano/Plutus) architectures through side-by-side implementations, showing how paradigms induce different error classes, though their analysis was theoretical. Bartoletti et al.~\cite{bartoletti2024smart} evaluated multiple languages using a benchmark suite, but conclusions were feature-based rather than behaviorally measured.

For Move specifically, Song et al.~\cite{song2024empirical} analyzed tens of thousands of contracts finding fewer asset-related bugs than Solidity, Giatzis et al.~\cite{giatzis2025comparative} proposed migration frameworks, and Bartoletti et al.~\cite{bartoletti2025formal} compared verification tooling, but none empirically measured developer security practices through controlled comparison.

Coblenz et al.~\cite{coblenz2020can} showed empirically that ownership-oriented type systems reduce bug classes while remaining usable, comparing Obsidian with Solidity in a controlled study.
Our study extends this methodology to Move using contract pairs and a set of metrics, while adding developer perception data absent from prior work.
\vspace{-5pt}
\section{Dataset and Methodology}

We analyzed smart contracts from the GitHub Rosetta Smart Contracts repository~\cite{rosetta2024}, a collection of use case implementations across multiple blockchain platforms and programming languages, designed to enable controlled comparisons by having the same development team implement identical use cases in different languages.

We selected all contracts with complete implementations in both Solidity (Ethereum) and Move (Sui). At the time of analysis (January 2025), this yielded 24 contracts (12 pairs) spanning different smart contract patterns (transfers, storage), financial primitives (auctions, escrow, vesting), and complex protocols (HTLC, oracles). The diversity ensures findings are not specific to a narrow application domain. Since all implementations were created by the same team, confounding factors related to programming skill, coding style, experience, and temporal language evolution are controlled. The complete repository is publicly available on GitHub, enabling independent verification.

We employed a mixed-methods approach combining quantitative code analysis (RQ1) with qualitative developer surveys (RQ2) to triangulate objective metrics with subjective perceptions.

\paragraph{Code Characteristics and Security Overhead (RQ1)}

We extracted code characteristics from all 24 contracts using regular expression pattern matching. For each contract, we extracted nine metrics. For size and structure we count \textbf{Source Lines of Code (SLOC)} considering non-blank, non-comment lines~\cite{jones2007estimating}. SLOC provides a language-neutral measure of verbosity. Then we capture modularity considering \textbf{Function Count}.
In Move, this includes \texttt{public} and \texttt{entry} modifiers. \textbf{Maximum Nesting} approximates cognitive complexity:

\begin{equation}
\text{MaxNesting} = \max_{i \in \text{lines}} \{\text{depth}(i)\}
\end{equation}
where $\text{depth}(i)$ counts open braces at line $i$.

For complexity, following McCabe~\cite{mccabe1976complexity}, we consider \textbf{Cyclomatic Complexity} counting linearly independent paths: $CC = 1 + \sum_{i=1}^{n} d_i$, 
where $d_i$ represents decision points (if, while, for, case, \&\&, $||$).

Then \textbf{Security Checks} counts explicit validation statements:
\texttt{require()} and \texttt{assert()} in Solidity and \texttt{assert!()} in Move. Insufficient validation is a primary attack vector~\cite{atzei2017survey,chen2020survey}. \textbf{Security Density} normalizes by code size:
\begin{equation}
\text{SecurityDensity} = \frac{\text{SecurityChecks}}{\text{SLOC}} \times 100
\end{equation}
This is our primary dependent variable, directly quantifying security overhead across contracts of different sizes. \textbf{External Calls} (Solidity only) counts low-level interactions (\text{call}, \text{delegatecall}, \text{staticcall}).
These introduce reentrancy risk~\cite{chen2020survey}; Move's type system prohibits such calls by design. \textbf{Type Annotations} (Move only) counts generic type parameters:
\begin{equation}
\text{TypeAnnotations} = |\{{\small \langle T\rangle, \langle T\text{:key}\rangle, \langle T\text{:store}\rangle, \text{etc.}}\}|
\end{equation}
These enforce compile-time safety but increase syntactic overhead. \textbf{Resource Operations} (Move only) counts ownership-manipulating calls:
\begin{equation}
\text{ResourceOps} = |\{{\small \text{transfer}, \text{move\_to}, \text{move\_from}, \text{borrow}, \ldots}\}|
\end{equation}
These replace implicit state management in Solidity and characterize Move's resource-oriented design.

All metrics are implemented in Python using regular expressions and validated against manual counts. Two independent raters manually analyzed three contract pairs (EditableNFT, HTLC, SimpleWallet; $n$=6 contracts). 
Inter-rater reliability reached mean Pearson $r = 0.968$ (5 of 7 metrics at $r > 0.95$), with automated vs. manual agreement at mean $r = 0.944$ and $r = 1.000$ for Security Checks and Function Count. The complete extraction code and validation data are available in our replication package.



Since each pair represents the same functionality in two languages, we used paired statistical tests to isolate paradigm effects from use case complexity.

Given $n$=12 pairs, we employed Wilcoxon signed-rank tests (two-tailed, $\alpha=0.05$) as primary analysis, following recommendations for small samples in empirical software engineering~\cite{arcuri2011practical}. To control the false discovery rate across nine metrics, we applied Benjamini-Hochberg FDR correction~\cite{benjamini1995controlling}, reporting both raw and adjusted p-values. For practical significance, we calculated Cohen's $d = \bar{D}/s_D$~\cite{cohen1988statistical}, interpreted as negligible ($|d|<0.2$), small ($0.2$--$0.5$), medium ($0.5$--$0.8$), or large ($|d|\geq0.8$).

\paragraph{Developer Survey (RQ2)}

We recruited 11 developers (Master's and PhD students) through a blockchain education program providing intensive training in both Solidity and Move, including lectures, hands-on development sessions, and a final hackathon. All participants had implemented contracts in both languages, ensuring practical rather than superficial exposure.

The survey comprised three sections: \textbf{(1) Background} (2 items) captured experience levels using categorical scales; \textbf{(2) Quantitative perceptions} (12 items) used seven-point Likert scales 
measuring learning and transition difficulty, documentation quality, tutorial availability, tool quality, type system perception, development time, and safety confidence; \textbf{(3) Qualitative insights} (3 items) asked about preferred application domains, biggest challenges, and language preference with reasoning. Responses were anonymous; the instrument is included in the replication package.

For paired comparisons (documentation, tutorials, tools, learning difficulty), we applied Wilcoxon signed-rank tests with Benjamini-Hochberg FDR correction at $\alpha=0.05$. For unipolar scales (type system, development time, safety confidence), we provide descriptive statistics only. Open-ended responses were analyzed using qualitative content analysis~\cite{braun2006thematic}, applying a structured coding scheme to identify recurring patterns across responses. Two independent coders achieved mean Cohen's $\kappa = 0.859$ (range 0.621--1.000; 93.2\% agreement across 44 coding decisions), confirming the validity of the coding scheme.
\vspace{-6pt}
\section{RQ1: Code Characteristics and Security Overhead}

We analyzed 12 functionally-equivalent contract pairs from the Rosetta repository, comparing code characteristics between Solidity (imperative) and Move (resource-oriented) implementations. All statistical comparisons used the Wilcoxon signed-rank test ($\alpha=0.05$) with Benjamini-Hochberg FDR correction for multiple comparisons, appropriate for small paired samples. Table~\ref{tab:rq1_summary} presents the complete results.

\vspace{-5pt}

\paragraph{Security Overhead} 
Move contracts required significantly fewer explicit security checks than Solidity contracts. Security check density—defined as the percentage of code lines dedicated to security validations (require/assert statements)—was 6.7\% in Move versus 16.8\% in Solidity (Wilcoxon $p=0.002$, Cohen's $d=-1.75$), representing a 60\% reduction. In absolute terms, Move contracts averaged 3.8 security checks compared to 6.0 in Solidity ($p=0.003$, $d=-1.40$). 
Solidity contracts employed low-level external calls (mean=1.3 per contract), while Move contracts used none ($p=0.005$, $d=-1.35$). Move's paradigm prohibits such calls, eliminating an entire class of vulnerabilities.

\vspace{-5pt}
\paragraph{Code Verbosity}
Move contracts were larger than Solidity contracts: 49.8 SLOC versus 33.9 SLOC (mean), a 47\% increase ($p=0.002$, $d=1.90$).
Type annotations appeared an average of 12.1 times per Move contract versus zero in Solidity ($p=0.002$, $d=2.90$). Similarly, resource-specific operations (coin transfers, dynamic field manipulations) averaged 7.8 per contract in Move versus zero in Solidity ($p=0.002$, $d=1.79$). While Move's median (54 SLOC) exceeds Solidity's (36.5 SLOC), both languages show similar variance, indicating the size increase is consistent across contract types rather than specific to certain use cases.


\vspace{-5pt}
\paragraph{Logical Complexity}
Despite differences in code size and paradigm, cyclomatic complexity was equivalent between languages: median=1.0 for both (mean=1.75, $p=1.0$, $d=0.00$). 
This finding confirms that the contracts implement equivalent logic—differences arise from paradigm design rather than algorithmic complexity. Maximum nesting depth showed no significant difference (Move: 3.0, Solidity: 2.8, $p=0.703$, $d=0.29$), further supporting functional equivalence.

\begin{figure}[H]
\centering
\includegraphics[width=\linewidth]{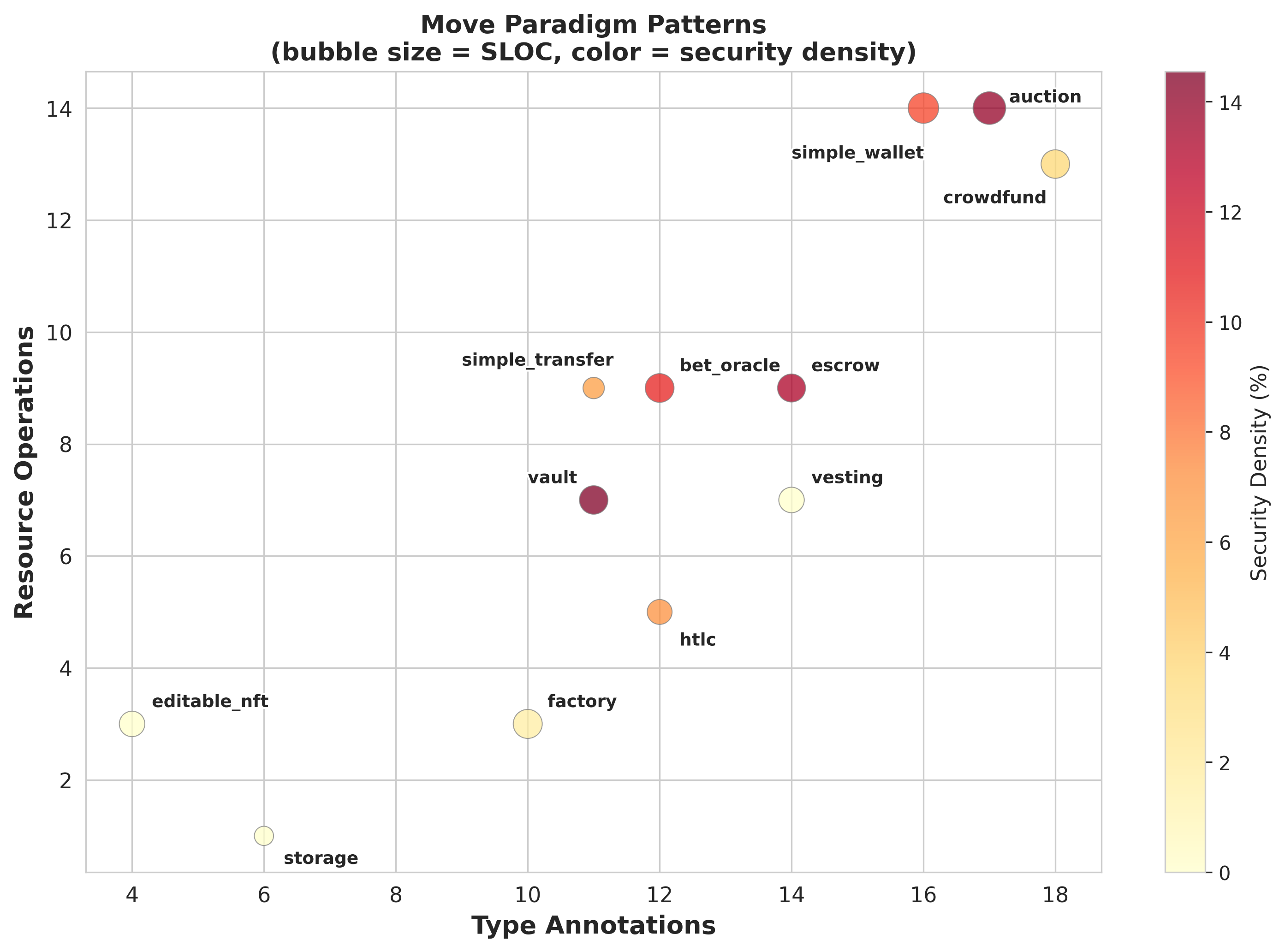}
\caption{Resource-oriented patterns in Move contracts. Bubble size represents SLOC; color represents security density. Type annotations and resource operations characterize Move's paradigm.}
\label{fig:paradigm_patterns}
\end{figure}

\vspace{-5pt}
\paragraph{Paradigm-Specific Patterns}
Move contracts exhibited clear resource-oriented design patterns. Figure~\ref{fig:paradigm_patterns} visualizes the relationship between type annotations and resource operations in Move contracts. Larger contracts (bubble size indicates SLOC) tend to have more type annotations, while security density (color) varies independently—some large contracts (e.g., crowdfund, auction) maintain higher security density despite the type system, while others (editable\_nft, vesting) rely entirely on type safety.


From Table~\ref{tab:rq1_summary} seven of nine metrics showed significant differences with large effect sizes. The two non-significant metrics—cyclomatic complexity and max nesting depth—are precisely those that should remain equivalent for functionally-identical implementations.

\begin{table}[h]
\centering
\caption{Code characteristics comparison (n=12 contract pairs). All tests: Wilcoxon signed-rank with FDR correction, $\alpha=0.05$. P-values are FDR-adjusted.}
\label{tab:rq1_summary}
\resizebox{\linewidth}{!}{
\begin{tabular}{lrrrrr}
\toprule
\textbf{Metric} & \textbf{Solidity} & \textbf{Move} & \textbf{Diff} & \textbf{$p$-value} & \textbf{Effect} \\
 & \textbf{(M±SD)} & \textbf{(M±SD)} & \textbf{(\%)} & & \textbf{Size ($d$)} \\
\midrule
Security Density (\%) & 16.8±9.6 & 6.7±5.6 & -60\% & \textbf{0.002} & Large (-1.75) \\
Code Lines (SLOC) & 33.9±11.7 & 49.8±13.2 & +47\% & \textbf{0.002} & Large (1.90) \\
Cyclomatic Complexity & 1.75±1.1 & 1.75±1.0 & 0\% & 1.0 & None (0.00) \\
Security Checks (abs.) & 6.0±3.8 & 3.8±3.5 & -38\% & \textbf{0.003} & Large (-1.40) \\
Type Annotations & 0±0 & 12.1±4.2 & --- & \textbf{0.002} & Large (2.90) \\
Resource Operations & 0±0 & 7.8±4.4 & --- & \textbf{0.002} & Large (1.79) \\
External Calls & 1.3±1.0 & 0±0 & --- & \textbf{0.005} & Large (-1.35) \\
Function Count & 3.3±1.0 & 4.2±1.0 & +28\% & \textbf{0.002} & Large (3.18) \\
Max Nesting Depth & 2.8±0.6 & 3.0±0.0 & +6\% & 0.703 & Small (0.29) \\
\bottomrule
\end{tabular}
}
\end{table}

\vspace{-5pt}

\begin{tcolorbox}[right=0.1cm,left=0.1cm,top=0.1cm,bottom=0.1cm]
\textbf{Answer to RQ1}: Resource-oriented programming (Move) reduces explicit security overhead by 60\% compared to imperative programming (Solidity) ($p=0.002$, $d=-1.75$). This reduction comes at the cost of 47\% more code ($p=0.002$, $d=1.90$), primarily from explicit type annotations (12 per contract vs. 0) and resource operations (8 per contract vs. 0). Cyclomatic complexity remains identical ($p=1.0$), confirming that differences stem from paradigm design rather than algorithmic complexity. Move's type system shifts security from runtime checks to compile-time guarantees.
\end{tcolorbox}
\vspace{-6pt}
\section{RQ2: Developer Experience and Perceptions}

To complement the code analysis, we surveyed 11 developers (Master's and PhD students from a blockchain summer school) who had implemented contracts in both languages. Participants had varying experience levels: Solidity (Mean=1.8 years, SD=1.2) and Move (Mean=1.0 years, SD=0.8). We analyzed their responses using descriptive statistics for Likert-scale items and qualitative content analysis for open-ended questions.

\vspace{-4pt}

\paragraph{Learning Experience}

Developers rated Move as more difficult to learn (Median=5/7) compared to Solidity (Median=4/7), though this difference was not statistically significant (Wilcoxon $p=0.410$, $n=11$). The transition difficulty was rated as moderate (Median=4/7), with substantial individual variation (SD=1.3). Qualitative coding of challenges identified four recurring categories (C):

\textbf{C1: Paradigm Shift (3 respondents).} The most frequently reported challenge was adapting to Move's resource-oriented model. Respondents described this as requiring conceptual adjustment:

\textit{``The biggest challenge was adapting to Move's strict resource-oriented model after being used to Solidity's more flexible state and memory handling.''} (R1)


\textit{``The linear type nature of Move objects takes a while to get used to, but it is more intuitive than Solidity.''} (R9)

\textbf{C2: Security Concerns in Solidity (4 respondents).} Developers frequently mentioned security as a Solidity challenge:


\textit{``Understanding which checks to implement in order to make the code as secure as possible against attacks.''} (R8)

This aligns with RQ1's finding that Solidity contracts require 60\% more security checks.

\textbf{C3: Documentation Gaps in Move (3 respondents).} Some developers noted Move's less mature ecosystem, such as:

\textit{``Documentation in some cases is high level, few tools for IDE, and sometimes not support from the community in move.''} (R10)


Quantitatively, tutorial availability differed significantly: Solidity (Median=6/7) versus Move (Median=4/7), Wilcoxon $p=0.023$. Documentation and tool quality showed similar trends but were not statistically significant ($p>0.05$).

\textbf{C4: Language Evolution (1 respondent).} One developer noted Solidity's rapid evolution as a challenge for staying current (R11).

\vspace{-8pt}

\paragraph{Safety Perceptions}

\begin{figure}[H]
\centering
\includegraphics[width=\linewidth]{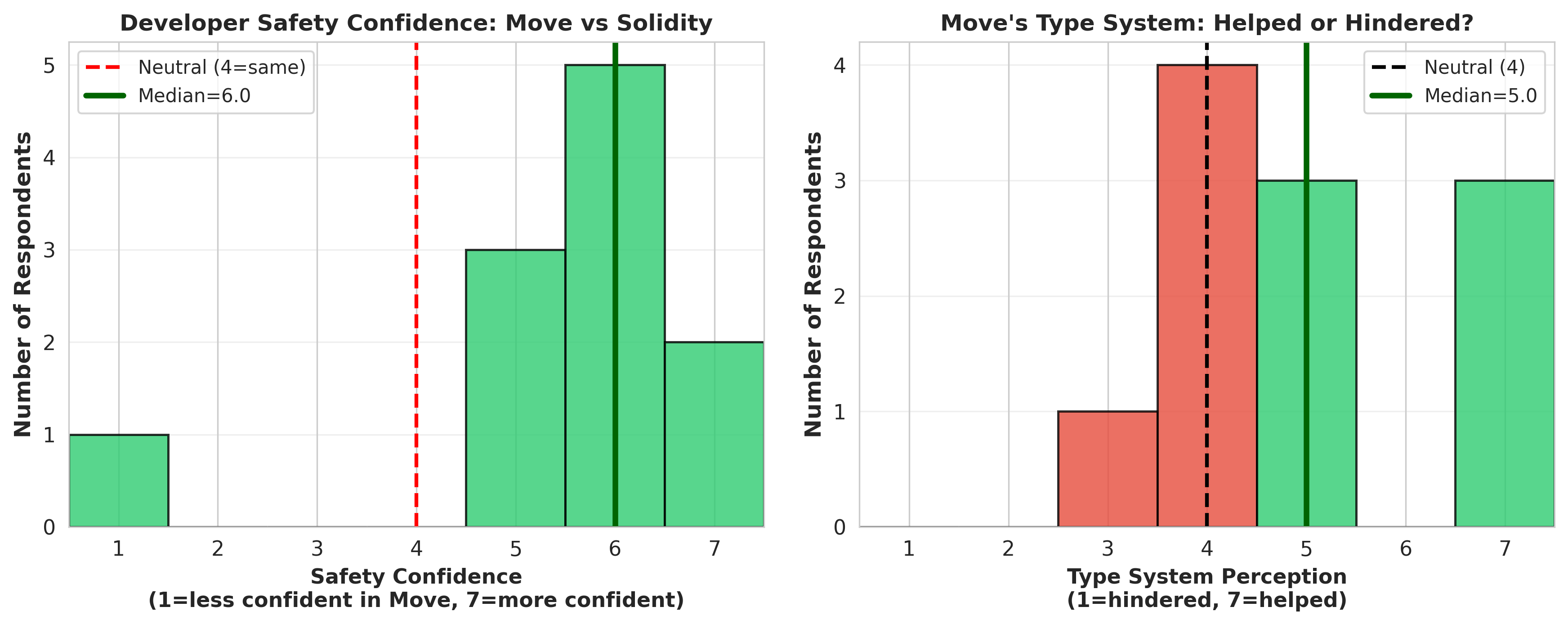}
\caption{Developer perceptions: (a) Safety confidence distribution showing most developers feel more confident in Move's safety (Median=6, n=10 above neutral), (b) Type system perception showing mixed but generally positive views (Median=5, n=6 found helpful).}
\label{fig:safety_perceptions}
\end{figure}

Despite learning challenges, developers expressed higher confidence in Move's safety (Median=6/7, where 7=much more confident in Move). Figure~\ref{fig:safety_perceptions}a shows the distribution: 10 of 11 respondents rated confidence above neutral (>4), with most clustering at 5--6. Only one respondent rated Move less safe (score=1).

Language preferences reflected this safety awareness. Six respondents (55\%) expressed preference for Move, citing safety explicitly:

\textit{``Yes, I would prefer using Move because its resource-oriented design provides stronger safety guarantees, reduces common smart contract vulnerabilities, and makes asset ownership and state changes more explicit and secure.''} (R1)


\textit{``For serious Blockchain applications where security is critical, Solidity should be avoided, given the plethora of better alternatives around today.''} (R9)

Two respondents (18\%) expressed conditional preferences based on use case, for example:

\textit{``It depends on the project, if it involves identities or objects and it is important to manage ownership, I would prefer move otherwise I would stick with solidity.''} (R7)

Three respondents (27\%) preferred their current language (Solidity) due to familiarity, comfort, or portability concerns.


\vspace{-8pt}

\paragraph{Type System Perceptions}

Developers' views on Move's type system were mixed but generally positive (Median=5/7). Figure~\ref{fig:safety_perceptions}b shows the distribution: 6 respondents found it helpful (>4), 4 rated it neutral (=4), and 1 found it hindering (<4). 
Development time perceptions showed that Move requires more time (Median=5/7, where 4=same time), consistent with the observed 47\% increase in code size from RQ1.

\vspace{-10pt}

\vspace{-15pt}
\paragraph{Application Contexts}

When asked where Move would be preferable, respondents most frequently selected: Identity and Governance (73\%, 8/11), Supply Chain Management (55\%, 6/11), DeFi applications (55\%, 6/11), Asset Tokenization (45\%, 5/11), NFTs and Gaming (45\%, 5/11). 

\begin{tcolorbox}[right=0.1cm,left=0.1cm,top=0.1cm,bottom=0.1cm]
\textbf{Answer to RQ2}: Developers experience the transition to resource-oriented programming as a moderate conceptual challenge (Median=5/7), primarily adapting to linear types and ownership models. Despite this learning curve and ecosystem maturity gaps (tutorials: Wilcoxon $p=0.023$), developers express higher safety confidence in Move (Median=6/7, 10 of 11 above neutral). Language preferences are context-dependent: 55\% prefer Move citing safety benefits, 27\% prefer Solidity for familiarity, and 18\% choose based on use case. Developers recognize Move as advantageous for ownership-critical domains (identity, asset management, DeFi), where the paradigm's safety guarantees justify the upfront learning investment.
\end{tcolorbox}
\section{Discussion}

This study provides preliminary evidence on how programming paradigms affect smart contract development through objective code analysis and developer experience. We discuss the alignment between these perspectives and practical implications.




A key finding is the alignment between objective metrics and developer perceptions: higher safety confidence (Median=6/7, 10 of 11 above neutral) corresponds to the 60\% reduction in security checks (p=0.002, d=-1.75), while verbosity reports (Median=5/7) align with the 47\% larger code size (p=0.002), with type annotations and resource operations materializing as the "additional syntax" developers described.

The identical cyclomatic complexity ($p=1.0$) validates the intuition that Move's learning curve stems from conceptual adaptation to ownership models rather than algorithmic complexity. As one respondent noted, linear types ``take a while to get used to, but [are] more intuitive than Solidity'' once understood.


Paradigm adaptation was the primary challenge (3 of 11 respondents), yet this barrier did not prevent safety recognition—55\% explicitly preferred Move for security. The tutorial availability gap (Solidity Median=6/7 vs. Move Median=4/7, $p=0.023$) represents a concrete adoption barrier; improved learning resources could lower this cost while preserving safety benefits.

Our data suggest this investment is most justified for ownership-critical domains: identity management (73\%), supply chain (55\%), and DeFi (55\%)—contexts where security failures carry high costs. For applications prioritizing development speed, team familiarity with Solidity, or access to mature tooling, imperative programming may remain preferable. These findings, while preliminary ($n$=12 pairs, 11 developers), suggest gradual adoption through pilot projects rather than wholesale migration.
\vspace{-10pt}
\section{Threats to Validity}

As a preliminary study, our findings are subject to limitations that must be considered when interpreting results.

\textit{Internal Validity}
Metric extraction relied on regular expression pattern matching, which could miss or misidentify language constructs. We mitigated this through manual validation by two independent raters on 25\% of contracts (3 pairs), achieving high inter-rater reliability (mean Pearson $r=0.968$) and strong automated-vs-manual agreement ($r=0.944$). Edge cases or unusual coding patterns may still introduce measurement error.

\textit{External Validity}
Our sample is small: 12 contract pairs and 11 survey respondents, limiting generalizability to larger codebases or broader developer populations. The survey recruited from a single blockchain education program; while this ensured hands-on experience with both languages, student participants may differ from professional developers with production experience.

\textit{Construct Validity}
Security check density captures explicit runtime validations but not all security dimensions—Move's type system may prevent vulnerabilities that Solidity addresses through mechanisms our metrics do not capture. Code size measures syntactic verbosity but not cognitive complexity or development effort. Survey responses on Likert scales are subject to response bias; we mitigated this through statistical tests and qualitative triangulation (content analysis coding: Cohen's $\kappa=0.859$).

\textit{Conclusion Validity}
Given $n=12$ pairs, statistical power is limited and some true differences may remain undetected. Large effect sizes (Cohen's $|d|>0.8$ for seven metrics) suggest findings are unlikely to be statistical artifacts, but replication with larger samples is necessary.
\vspace{-8pt}
\section{Conclusion}

This preliminary study compared imperative (Solidity) and resource-oriented (Move) smart contract paradigms through analysis of 12 functionally-equivalent contract pairs and a survey of 11 developers experienced in both languages. Our findings provide initial evidence that paradigm choice meaningfully affects both security overhead and developer experience.

The quantitative analysis revealed that Move's resource-oriented paradigm reduces explicit security overhead by 60\% compared to Solidity's imperative approach. This reduction comes at the cost of 47\% larger code size, primarily from type annotations and resource operations that enforce compile-time safety guarantees. Cyclomatic complexity remained identical between paradigms, confirming that observed differences stem from language design rather than algorithmic complexity.


Developer perceptions aligned with these objective metrics: despite rating Move as harder to learn and noting ecosystem gaps, most expressed higher safety confidence. Language preferences were context-dependent, with ownership-critical domains—identity management, asset tokenization, DeFi—identified as most suitable for resource-oriented programming.

For practitioners, resource-oriented paradigms are most advantageous when asset ownership is central and security requirements justify higher development time; transition costs recommend gradual adoption through pilot projects. 
 


Future work should extend these findings to larger contract repositories and more diverse developer populations; longitudinal studies tracking productivity and error rates, and empirical evaluation of actual vulnerability rates in deployed contracts, would complement these perceptions with behavioral outcomes.
\vspace{-9pt}

\begin{acks}
This work was partially supported by the SERICS project (PE00000014) under the MUR National Recovery and Resilience Plan funded by the European Union–NextGenerationEU, and by the AISAC project (MIMIT, ``Accordi per l'Innovazione'' 2021--2026, CUP: B29J23001120005, COR: 1607797).
\end{acks}
\vspace{-9pt}

\bibliographystyle{ACM-Reference-Format}
\bibliography{biblio}

\end{document}